# Density functional characterization of the antiferromagnetism in oxygen-deficient anatase and rutile TiO$_2$

Kesong Yang,[a,b] Ying Dai,[a]* Baibiao Huang,[a] and Yuanping Feng,[b†]

[a] *School of Physics, State Key Laboratory of Crystal Materials, Shandong University, Jinan 250100, People's Republic of China*

[b] *Department of Physics, National University of Singapore, 2 Science Drive 3, Singapore 117542*

We present theoretical evidence for local magnetic moments on Ti$^{3+}$ ions in oxygen-deficient anatase and rutile TiO$_2$ observed in a recent experiment [S. Zhou, et al., Phys. Rev. B **79**, 113201 (2009)]. Results of our first-principles GGA+U calculations reveal that an oxygen vacancy converts two Ti$^{4+}$ ions to two Ti$^{3+}$ ions in anatase phase, which results in a local magnetic moment of 1.0 $\mu_B$ per Ti$^{3+}$. The two Ti$^{3+}$ ions, however, form a stable antiferromagnetic state, and similar antiferromagnetism is also observed in oxygen-deficient rutile phase TiO$_2$. The calculated results are in good agreement with the experimentally observed antiferromagnetic-like behavior in oxygen-deficient Ti-O systems.

Owning to their promising applications in the spintronics, numerous attempts have been made to prepare diluted magnetic semiconductors (DMS) by doping semiconductors, particularly transition metal oxides $TiO_2$ and ZnO, with magnetic ions.[1, 2] Recently, high-temperature ferromagnetism was found in one class of semiconductors without magnetic ion dopants,[3, 4] which is referred to as the $d^0$ magnetism.[5] For example, ferromagnetism was observed in undoped $HfO_2$ consisting of nonmagnetic ions $Hf^{4+}$ ($d^0$) and $O^{2-}$,[4] for which electronic structure calculations showed that the local magnetic moments produced by the Hf vacancies are ferromagnetically coupled.[6] Room-temperature ferromagnetism was also reported in other undoped semiconductors such as $In_2O_3$, $SnO_2$, and $TiO_2$.[7-10] Among the various oxides, magnetic property of undoped $TiO_2$ has been widely studied.[9-12] However, despite of numerous studies, the origin of the ferromagnetism in undoped $TiO_2$ remains unclear. Both oxygen vacancy and titanium vacancy were proposed to be responsible for the ferromagnetism. On one hand, theoretical studies indicated that the cation vacancy or divacancy are ferromagnetically coupled,[13, 14] similar to the case of undoped $HfO_2$.[6] But on the other hand, more and more experimental evidences show that the magnetic property of undoped $TiO_2$ is strongly related to oxygen vacancy, and thus it was thought to be the source of room-temperature ferromagnetism in undoped semiconducting or insulating oxides.[7, 10-12] Although the electronic state induced by the oxygen vacancy has been studied by first-principles theoretical calculations,[15, 16] few works have been focused on its magnetic property. In particular, a recent experiment reported the presence of $Ti^{3+}$ ions in rutile phase at the substitutional sites near oxygen vacancies, and the unpaired $3d$ electron of the $Ti^{3+}$ ($d^1$) ion provides the local magnetic moment.[17] Consequently, an interesting question occurs to us:

how do the local magnetic moments of the $Ti^{3+}$ ions interact, and whether the coupling is ferromagnetic or antiferromagnetic. Since oxygen vacancies are very common in oxides, it would be useful to clarify the influence of oxygen vacancy on the magnetic property of undoped $TiO_2$.

In this work, we investigate magnetic property of oxygen-deficient anatase and rutile $TiO_2$ by first-principles GGA+U electronic structure calculations. We find that owning to the charge imbalance created by the oxygen vacancy, two excess electrons occupy the localized 3d orbitals of the nearest neighbor Ti, thereby converting two $Ti^{4+}$ ions to two $Ti^{3+}$ ions in anatase phase, each with a local magnetic moment of 1.0 $\mu_B$. However, the two $Ti^{3+}$ ions form a stable antiferromagnetic configuration, and similar antiferromagnetism is also found in rutile $TiO_2$.

Our spin-polarized GGA+U electronic structure calculations are carried out using the Vienna *ab inito* simulation package.[18, 19] Oxygen-deficient anatase and rutile models are constructed by removing an oxygen atom from 48-atom $2 \times 2 \times 1$ anatase supercell and 72-atom $2 \times 2 \times 3$ rutile supercell, respectively. Projector augmented wave (PAW) potentials are used to describe the electron-ion interaction while the generalized gradient approximation (GGA) parameterized by Perdew and Wang (PW91) is used for electron exchange-correlation functional.[20] The cut-off energy of 400 eV and a 2×2×2 k-point set centered at Γ point are sufficient to converge the total energy to within a tolerance of $10^{-6}$ eV. The lattice parameters and all the atomic positions are fully optimized until all components of the residual forces are smaller than 0.01 eV/Å. In our GGA+U calculations, the on-site effective U parameter ($U_{eff}$=U-J = 5.8 eV) proposed by Dudarev *et al.*[21] is adopted for Ti 3d electron,[22] which is in agreement with the optimal U value (5.5±0.5 eV).[23]

The total density of states (TDOS) plots of oxygen-deficient anatase and rutile $TiO_2$ are presented in **Fig. 1** and **Fig. 2,** respectively. For oxygen-deficient anatase phase, the calculated results show that the TDOS is spin-unpolarized, and some defect states are localized in the band gap. Interestingly, the partial density of states (PDOS) of the two Ti ions around the oxygen vacancy given in **Figs. 1b** and **1c,** respectively, are spin polarized. However, their magnetic moments are in opposite directions, which results in a zero total magnetic moment. In contrast, the PDOS of the third nearest neighbor Ti ion of the oxygen vacancy does not show any spin polarization, as shown by the following spin density distribution of **Fig. 4a**. This indicates that the two electrons introduced by the neutral oxygen vacancy are captured by the two neighboring Ti ions, forming two $T^{3+}$ ions with a local spin magnetic moment of 1.0 $\mu_B$. This, however, is in contrast to results of a recent theoretical study based on the local spin density approximation (LSDA), in which the authors suggested that the oxygen vacancy does not produce any magnetic moments.[14] This discrepancy could be due to the fact that the standard DFT calculations in the scheme of either local density approximation (LDA) or generalized gradient approximation (GGA) cannot treat properly the strong Coulomb interaction between 3$d$ electrons, and thus may lead to an inadequate description of 3d states of $Ti^{3+}$ in the oxygen-deficient $TiO_2$ system. As in the case of oxygen-deficient anatase $TiO_2$, some localized band gap states introduced by the oxygen vacancy are also found in oxygen-deficient rutile phase of $TiO_2$ through GGA+U calculations (see **Fig. 2**), and the further PDOS (**Figs. 2b** and **2c**) indicates that these localized impurity states consist of the spin-polarized states of Ti ions around the oxygen vacancy. In rutile phase, in contrast with anatase phase, the electronic states of the three Ti ions around the oxygen vacancy are both spin-polarized, and the calculated PDOS for two equivalent Ti ions is down-spin, and the third one is up-spin, which leads to a total spin magnetic moment of zero. This result can be clearly reflected by the following spin density distribution plot of **Fig. 4b.** This suggests that one $Ti^{4+}$ ion is reduced to a $Ti^{3+}$ with a spin magnetic moment of 1.0 $\mu_B$, while the other two $Ti^{4+}$ ions share one remaining electron introduced by oxygen vacancy and hence are reduced into low-state Ti ions (close to +3.5), respectively. The different electron states of these Ti ions also lead to different impurity level positions in the band gap, as shown in **Fig. 2**. The different electronic distribution on the adjacent Ti ions around the oxygen vacancy in oxygen-deficient anatase and rutile $TiO_2$ can be explained by their different local geometrical structures. In **Fig. 3**, we plot the local structures of oxygen-deficient anatase and rutile $TiO_2$ models, respectively. In anatase phase, after geometrical optimization, the distance between the two equivalent Ti ions and the oxygen vacancy (marked as yellow color in **Fig. 3a** and labeled as Ti-$V_O$ bond in the following discussion for convenience) are smaller than the third Ti-$V_O$ bond (marked as green color in **Fig. 3a**) (1.950 vs. 2.03 Å), and thus the two electrons introduced by the oxygen vacancy will occupy the two equivalent Ti ions preferredly, producing a magnetic moment of 1.0 $\mu_B$ on each Ti ion with opposite spin directions. In contrast, in rutile phase, upon structural relaxation, two equivalent Ti-$V_O$ bonds (marked as yellow color in **Fig. 3b**) are longer than the third Ti-$V_O$ bond (marked as green color in **Fig. 3b**). As a result, one of the two electrons induced by the oxygen vacancy will firstly occupy the nearest Ti ion, which reduces the $Ti^{4+}$ to $Ti^{3+}$ with a spin magnetic moment of 1.0 $\mu_B$, and the other electron was captured by the two equivalent $Ti^{4+}$ ions and forms a total magnetic moment of 1.0 $\mu_B$ with opposite spin direction. In conclusion, these calculated results provide a clear theoretical evidence for experimentally observed $Ti^{3+}$ ions in oxygen-deficient rutile $TiO_2$.[17]

To further investigate the magnetic coupling characteristic between the local moments from the induced $Ti^{3+}$ ions by oxygen vacancy in anatase and rutile $TiO_2$, we compared the total energies of ferromagnetic and antiferromagnetic alignments of the magnetic moments on the generated $Ti^{3+}$ ions. It is found that the antiferromagnetic state is more stable than the ferromagnetic state by 474 meV for anatase phase $TiO_2$, and by 175 meV for rutile phase $TiO_2$. For anatase phase, we repeated our calculation using a 96-atom 2 × 2 × 2 supercell in which two oxygen vacancies were separated by the distance about 7.6 Å. The antiferromagnetic alignment of the two $Ti^{3+}$ ions around each oxygen vacancy remains more energetically favorable compared to the ferromagnetic alignment. The calculated results are in good agreement with the experimentally observed antiferromagnetic-like behavior in oxygen-deficient Ti-O system.[24]

To understand the nature of spin exchange coupling in oxygen-deficient anatase and rutile $TiO_2$, we show their spin density distributions under antiferromagnetic alignment in **Figs. 4a** and **4b**, respectively. In anatase phase, the spin density is mainly distributed on the two $Ti^{3+}$ ions and they have opposite spin directions. In contrast, in rutile phase, the two equivalent Ti ions share one electron introduced by oxygen vacancy and thus they have the same spin direction, and the third $Ti^{3+}$ ($d^1$) ion possesses the one remaining electron, which provides a magnetic moment of 1.0 $\mu_B$ with opposite spin direction. It is well-known that the superexchange model was widely used to explain the antiferromagnetic coupling between two next-nearest neighbor cations through a non-magnetic anion (MnO, FeO, etc.).[25] However, in the oxygen-deficient $TiO_2$ model, the middle non-magnetic oxygen atom is removed, and thus the classical superexchange mechanism is not appropriate to explain the antiferromagnetism in this model. Therefore, in light of the information obtained from the first-principles calculations, we propose another possible superexchange model on the basis of the indirect *d-d* hopping between the paired $Ti^{3+}$ ions via the adjacent $Ti^{4+}$, as illustrated in **Fig. 5,** to explain the AFM coupling in oxygen-deficient $TiO_2$. The electron in each $Ti^{3+}$ ($d^1$) ion hops to the *d* orbital of the adjacent $Ti^{4+}$ through the oxygen ions, making the $Ti^{4+}$ ion a low-spin $Ti^{2+}$ ($d^2$) state. In this process, the electrons do not have to change their spin directions, and thus the overall energy saving can lead to an antiferromagnetic alignment of the two $Ti^{3+}$ ions.

As a comparison, we also investigated the electronic and magnetic property of titanium-deficient anatase $TiO_2$. Similar to the case of $HfO_2$ [6] and CaO[26], the presence of cation vacancy causes a clear spin split in the valence band, and a total magnetic moment of 4.0 $\mu_B$ was produced, mainly contributed by the six adjacent oxygen ions around the titanium vacancy which is consistent with results of previous calculations based on LSDA or GGA functional.[13, 14] Further calculations were carried out to assess the relative stability of the ferromagnetic and antiferromagnetic alignments between the magnetic moments localized on different titanium vacancies using the 96-atom $2 \times 2 \times 2$ supercell where the distance between two titanium vacancies is 7.6 Å, and the FM state is found to be more stable than the AFM state by 112 meV, indicating a substantially long-range ferromagnetic ordering of cation vacancy induced by local magnetic moments in titanium-deficient $TiO_2$. As a result, it is proposed that the carriers, i.e., holes from the *p* orbitals of these oxygen ions, are thought to induce the long-range ferromagnetism, and similar ferromagnetically coupled state is also expected in other cation-deficient semiconductors such as $In_2O_3$, $SnO_2$, and CdS.

In summary, first-principles GGA+U electronic structure calculations are carried out to investigate the magnetic property of oxygen-deficient anatase and rutile $TiO_2$. Results of the calculations show that excess electrons introduced by an oxygen vacancy convert two $Ti^{4+}$ ions into two $Ti^{3+}$ ions and result in a local magnetic moment of about 1.0 $\mu_B$ per $Ti^{3+}$ ion in anatase phase. However, the two $Ti^{3+}$ ions form a stable antiferromagnetic state. Similar antiferromagnetism also appears in oxygen-deficient rutile phase $TiO_2$. The calculated results are consistent with the experimentally observed antiferromagnetic behavior in oxygen-deficient Ti-O system.

## ACKNOWLEDGMENTS

This work is supported by the Singapore National Research Foundation Competitive Research Program (Grant No. NRF-G-CRP 2007-05), National Basic Research Program of China (973 program, Grant No. 2007CB613302), National Natural Science Foundation of China under Grant No. 10774091, Natural Science Foundation of Shandong Province under Grant No. Y2007A18, and the Specialized Research Fund for the Doctoral Program of Higher Education 20060422023.

E-mail addresses: daiy60@sina.com (Ying Dai[*]), phyfyp@nus.edu.sg (Yuanping Feng[†])

**Figure Captions**

**Figure 1** Total (a) and partial (b, and c) DOS plots of oxygen-deficient anatase TiO2. The vertical dotted line indicates the Fermi level.

**Figure 2** Total (a) and partial (b, and c) DOS plots of oxygen-deficient rutile TiO2. The vertical dotted line indicates the Fermi level.

**Figure 3** Partial geometrical structures of the models for oxygen-deficient (a) anatase and (b) rutile $TiO_2$. The larger gray and small red spheres represent the Ti and O atoms respectively, and the white spheres show the position of oxygen vacancies.

**Figure 4** Spin densities around the oxygen vacancy in anatase (a) and rutile (b) $TiO_2$ under antiferromagnetic alignment. Yellow and cyan isosurfaces correspond to up- and down-spin densities, respectively.

**Figure 5** Schematic of the AFM coupling between the two $Ti^{3+}$ ions around the oxygen vacancy.

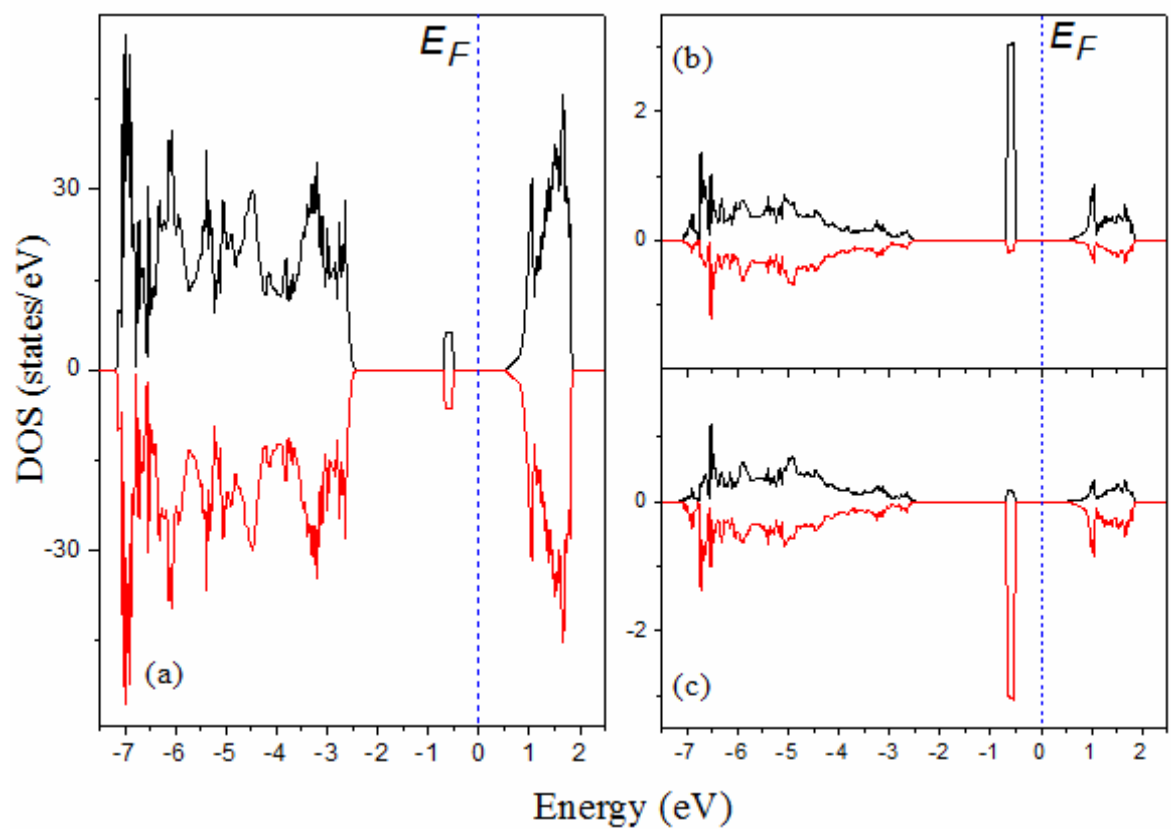

**Figure 1**

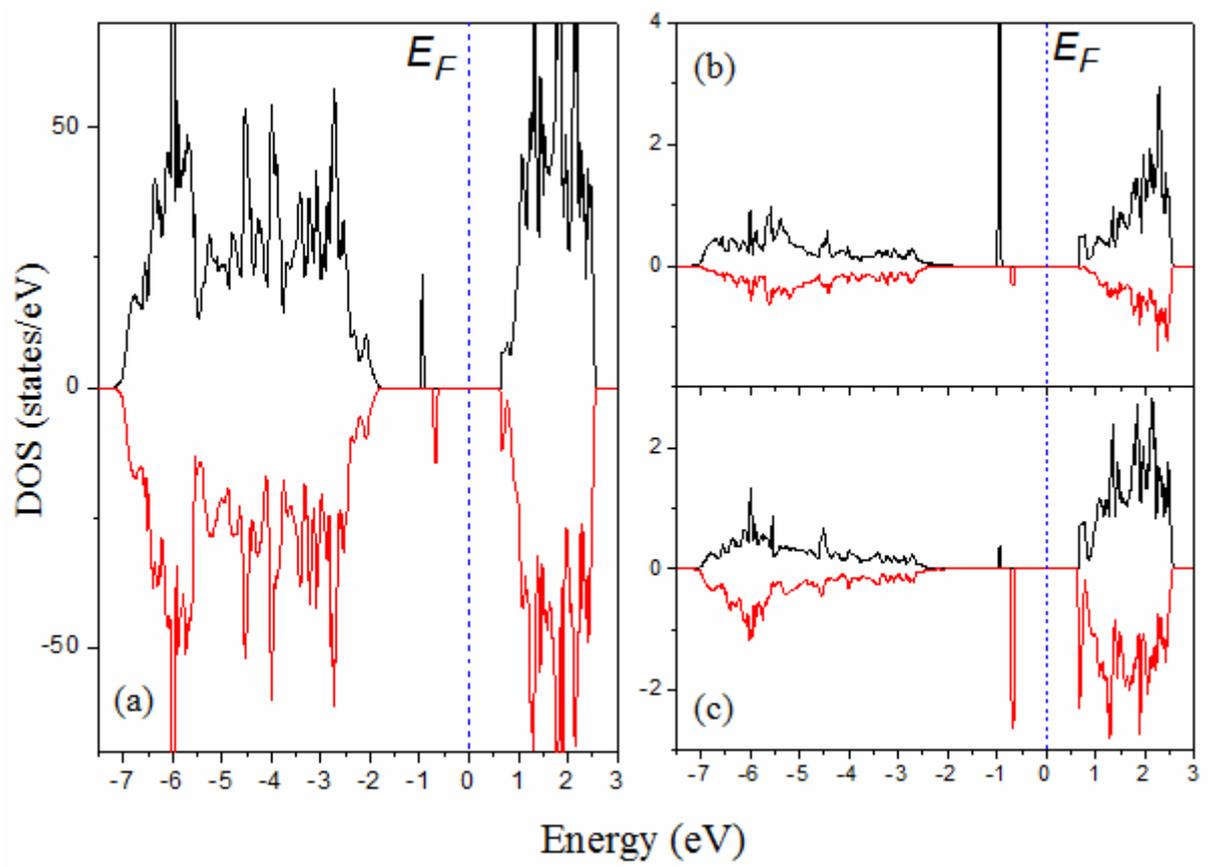

**Figure 2**

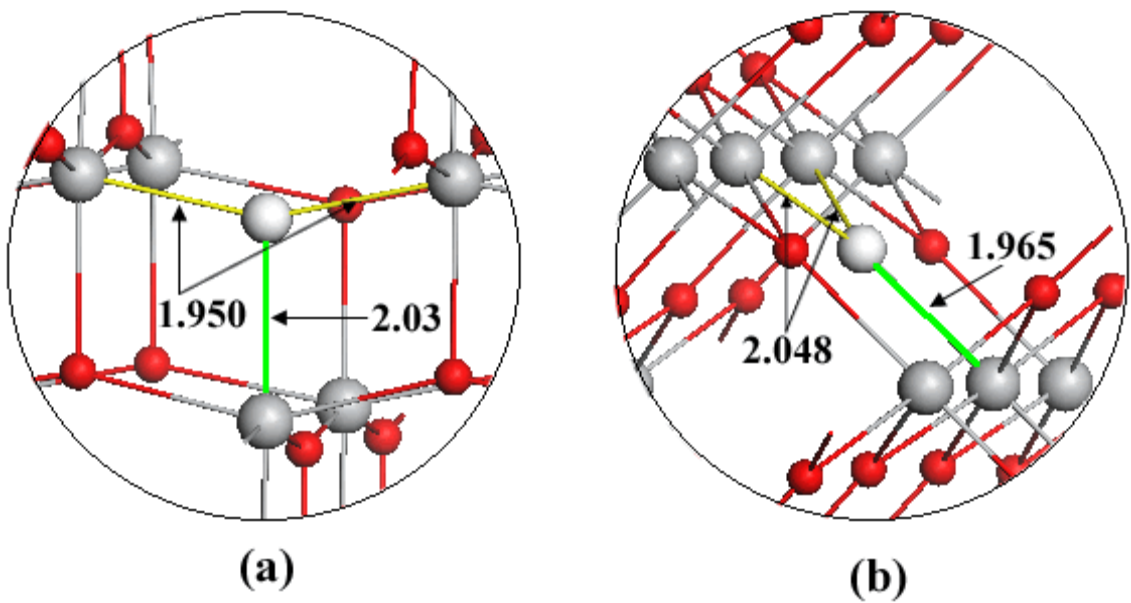

**Figure 3**

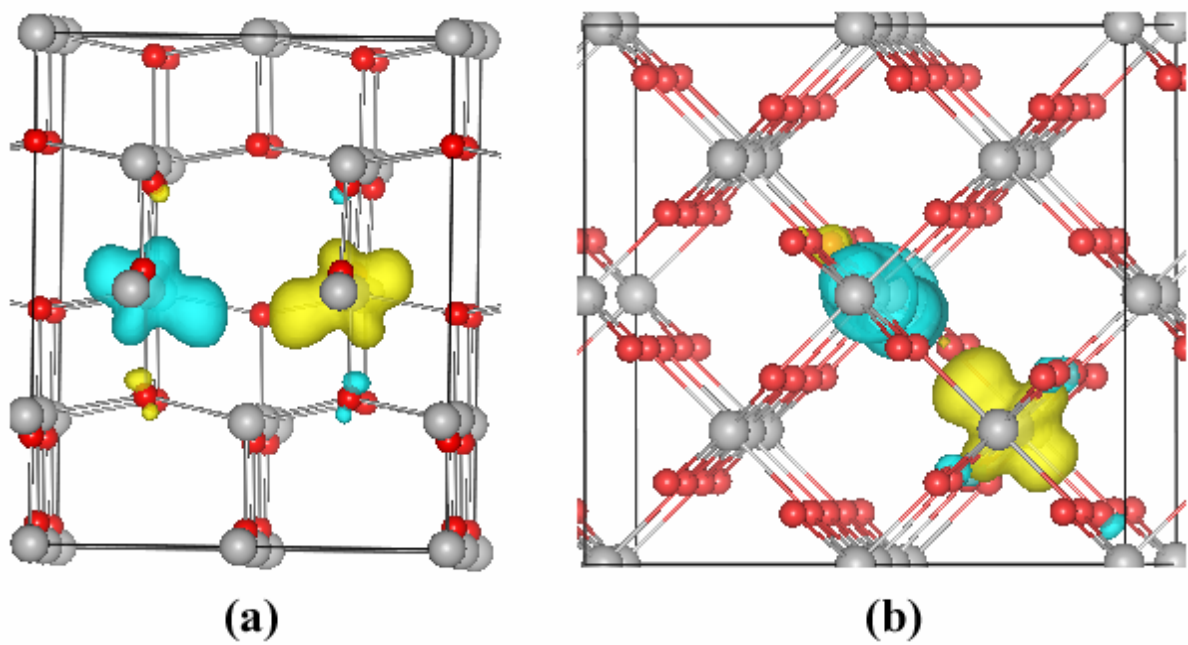

**Figure 4**

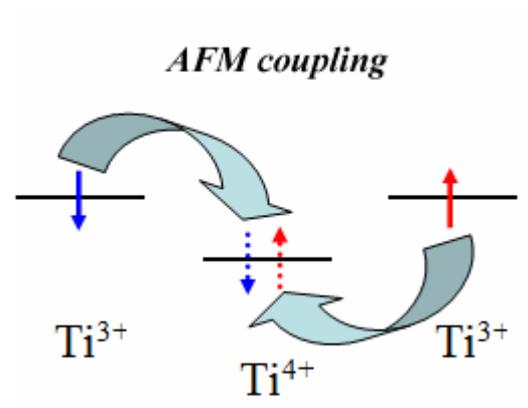

**Figure 5**